\documentclass{PHYEAUTH}

\usepackage{amsmath,graphicx}

\begin{document}

\begin{frontmatter}


\title{Determining Carrier Densities in InMnAs by Cyclotron Resonance}

\author[uf]{G. D. Sanders},
\author[uf]{Y. Sun},
\author[uf]{C. J. Stanton},
\author[rice]{G. A. Khodaparast},
\author[rice]{J. Kono},
\author[stanford]{D. S. King},
\author[okayama]{Y. H. Matsuda},
\author[ut]{S. Ikeda},
\author[ut]{N. Miura},
\author[tit]{A. Oiwa}, and
\author[tit]{H. Munekata}

\address[uf]{Department of Physics, University of Florida}
\address[rice]{Department of Electrical and Computer Engineering,
               Rice University}
\address[stanford]{Department of Applied Physics, Stanford University}
\address[okayama]{Department of Physics, Okayama University}
\address[ut]{Institute of Solid State Physics, University of Tokyo}
\address[tit]{Imaging Science and Engineering Lab,
              Tokyo Institute of Technology}

\begin{abstract}

Accurate determination of carrier densities in ferromagnetic
semiconductors by Hall measurements is hindered by the anomalous Hall
effect, and thus alternative methods are being sought. Here, we propose
that cyclotron resonance (CR) is an excellent method for carrier density
determination for InMnAs-based magnetic structures.  We develop a theory
for electronic and magneto-optical properties in narrow gap InMnAs films
and superlattices in ultrahigh magnetic fields oriented along [001].
In n-type InMnAs films and superlattices, we find that the e-active CR
peak field is pinned at low electron densities and then begins to shift
rapidly to higher fields above a critical electron concentration allowing
the electron density to be accurately calibrated.  In p-type InMnAs,
we observe two h-active CR peaks due to heavy and light holes. The
lineshapes depend on temperature and line broadening.  The light hole
CR requires higher hole densities and fields. Analyzing CR lineshapes
in p-films and superlattices can help determine hole densities.

\end{abstract}


\end{frontmatter}



\section{Introduction}

The determination of carrier densities in semiconductors is usually
carried out by Hall measurements.  In ferromagnetic semiconductors such
as InMnAs or GaMnAs however, the anomalous Hall effect below the Curie
temperature, $T_c$, can often times complicate the situation and make
the determination of carrier density difficult.

Recently,  several alternative methods for determining the carrier
densities  in ferromagnetic (III,Mn)V semiconductors have been proposed.
Some of these include: Raman Spectroscopy,\cite{Seoung} 
C-V measurements,\cite{Moryia} and  
infrared optical conductivity.\cite{Sinova}

In this paper, we propose that cylcotron resonance spectroscopy can
sometimes be used to very accurately determine the density for 
InMnAs-based magnetic structures. 

While the standard method for determining density from cyclotron
resonance is to integrate the total absorption, in the III-V dilute
magnetic semiconductors, owing to the large amount of magnetic
impurities, one must go to ultrahigh magnetic fields (40 T or more)
just to be able to observe the cyclotron resonance.  As a result,
there is sometimes a background signal which can influence the density
determined this way. Instead of using this standard method, we propose
an alternative scheme based on using information which can be obtained
from the CR resonant absorption such as the asymmetric lineshapes, and
the positions and relative intensities of different CR absorption
features.  We show that the lineshapes, postions, relative
intensities, and multiple transitions depend critically on the Fermi
energy and can thus allow one to accurately determine the carrier
density.


\begin{figure}[tbp]
\includegraphics[scale=0.48]{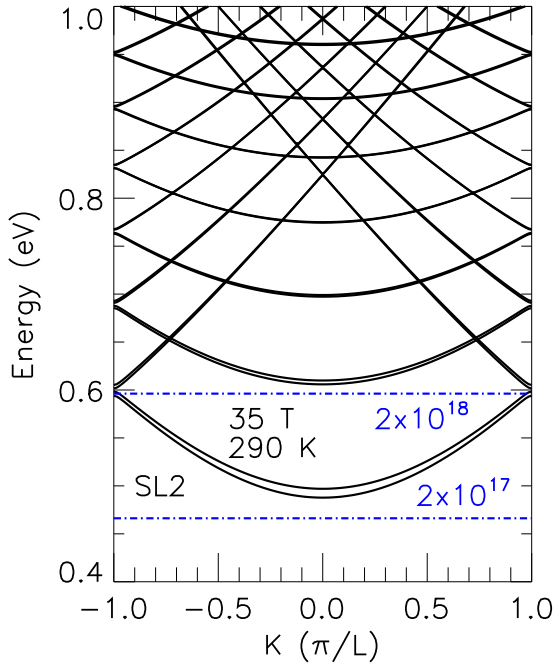}
\caption{Miniband structure for an n-type
InAs/In$_{0.894}$Mn$_{0.106}$As superlattice in a
35 Tesla field at 290 K. }
\label{fig1}
\end{figure}

\section{n-type results}

We first consider InAs which is periodically doped with Mn to produce
an n-type InAs/In$_{0.894}$Mn$_{0.106}$As superlattice
in which the InAs and InMnAs layers are each 50 \AA \ thick.
By periodically doping with Mn, we create a spin-dependent exchange
potential which tends to confine the spin-up electrons in the InAs
layer and spin-down electrons in the InMnAs layer.
We use a modified 8 band Pidgeon-Brown model for electronic and
optical properties in a high
magnetic field, $B \parallel \hat{z}$, where $\hat{z}$ is perpendicular
to the  superlattice layers.
The Pidgeon-Brown model \cite{pidgeon66} has been
generalized to include superlattice effects due to the periodic
Mn doping profile, the wavevector (k$_z$) dependence of the
electronic states, and the 
position dependent $s$-$d$ and $p$-$d$ exchange interactions
with localized Mn $d$-electrons. Magneto-optical properties and
cyclotron resonance are obtained using Fermi's golden rule to
compute the dielectric function. The Dirac delta functions appearing
in the golden rule are replaced by Lorentzian line shapes with the 
full width at half maximum (FWHM) being an input parameter. 
For details, see Ref. \cite{sanders03}.
The exchange interactions, which are parameterized by exchange
integrals $\alpha$ and $\beta$, are periodic with period
L = 100 \AA \ and give rise to minibands with
$\arrowvert k_z \arrowvert < \pi/L$.
The miniband structure for the superlattice at 
room temperature in a 35 Tesla field is shown in Fig.~\ref{fig1}.
The minibands are nearly free electron like and the zone 
folding of the Landau levels is clearly seen.
The Fermi levels for electron concentrations of $2 \times 10^{17}$ and
$2 \times 10^{18} \ \mbox{cm}^{-3}$ are shown in the figure.

\begin{figure}[tbp]
\includegraphics[scale=0.48]{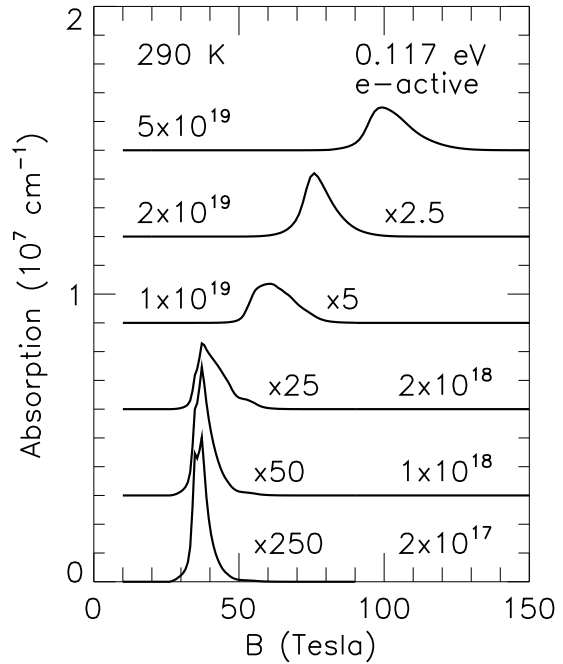}
\caption{Calculated room temperature CR for the n-type
InAs/In$_{0.894}$Mn$_{0.106}$As superlattice for different densities.}
\label{fig2}
\end{figure}

Room temperature e-active CR absorption vs.
electron density, $n$, are shown in Fig.~\ref{fig2} for
photon energies $\hbar \omega = 0.117 \ \mbox{eV}$ and $n$ ranging from
$2 \times 10^{17}$ to $5 \times 10^{19} \ \mbox{cm}^{-3}$.
The resonant field depends on carrier density. For
$n < 2 \times 10^{18} \ \mbox{cm}^{-3}$, the resonant field is
fixed around 35 Tesla and density determination based on the CR peak
position is impossible. The electronic band structure at the
resonant field is shown in Fig.~\ref{fig1}. At low carrier
concentrations, CR absorption occurs between the ground state and
first excited Landau levels.
For $n > 2 \times 10^{18} \ \mbox{cm}^{-3}$, this transition is 
suppressed by Pauli blocking as the first excited Landau levels start
to fill. This is evident from the position of the Fermi level
for $n = 2 \times 10^{18} \ \mbox{cm}^{-3}$ in Fig.~\ref{fig1}.
As $n$ continues to increase, the CR peak begins to
shift rapidly to higher fields making it possible for the electron
density to be accurately calibrated.
If $n = 10^{19} \mbox{cm}^{-3}$, one should be able to distinguish
between $1x10^{19} \mbox{cm}^{-3}$ and $1.1x10^{19} \mbox{cm}^{-3}$
based on the CR peak position for an accuracy of 10\% in the density.
For higher $n$, the accuracy should be even better as the CR peak
position shifts begins to shift more rapidly with $n$.

\begin{figure}[tbp]
\includegraphics[scale=0.42]{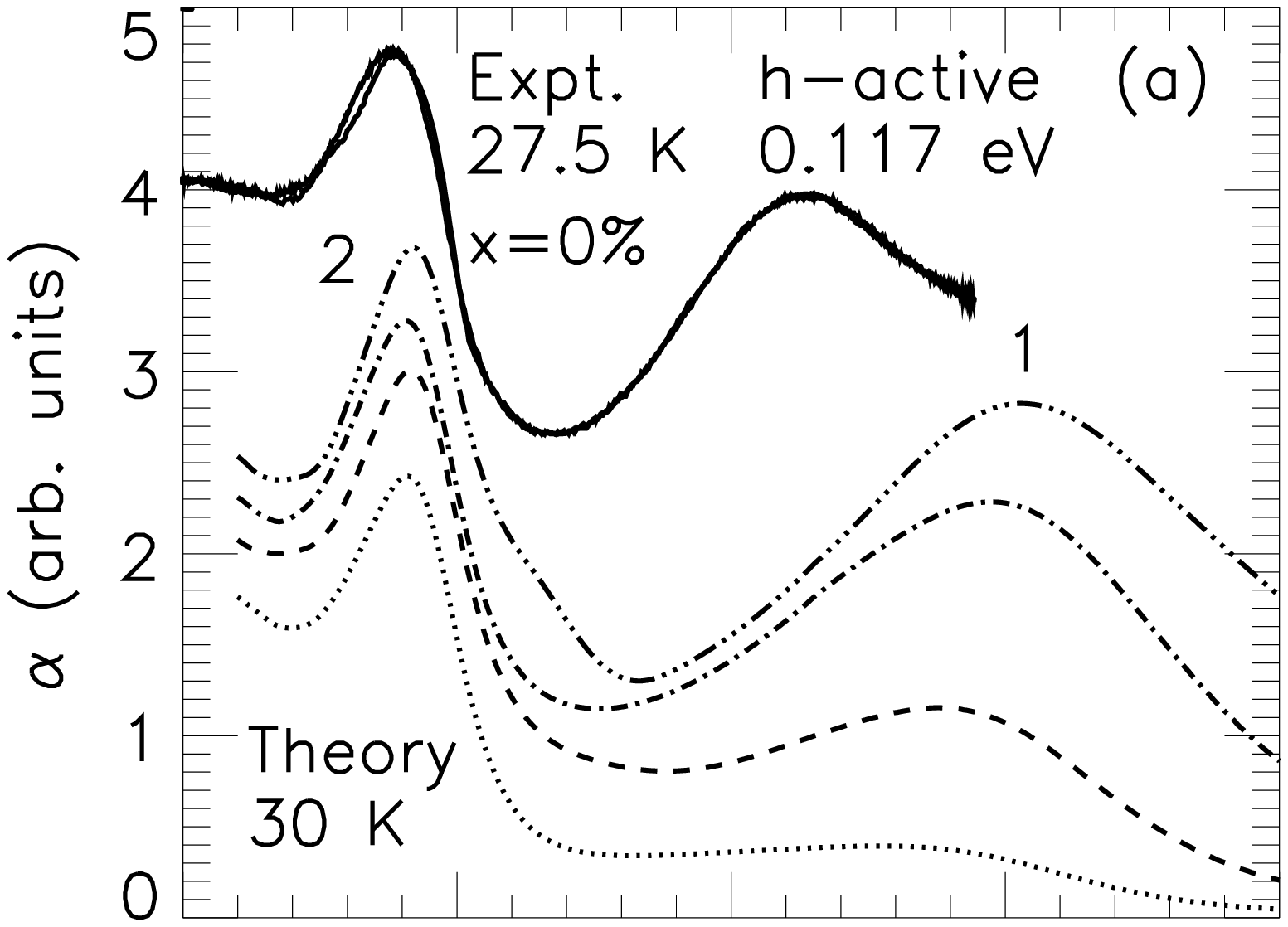}
\includegraphics[scale=0.42]{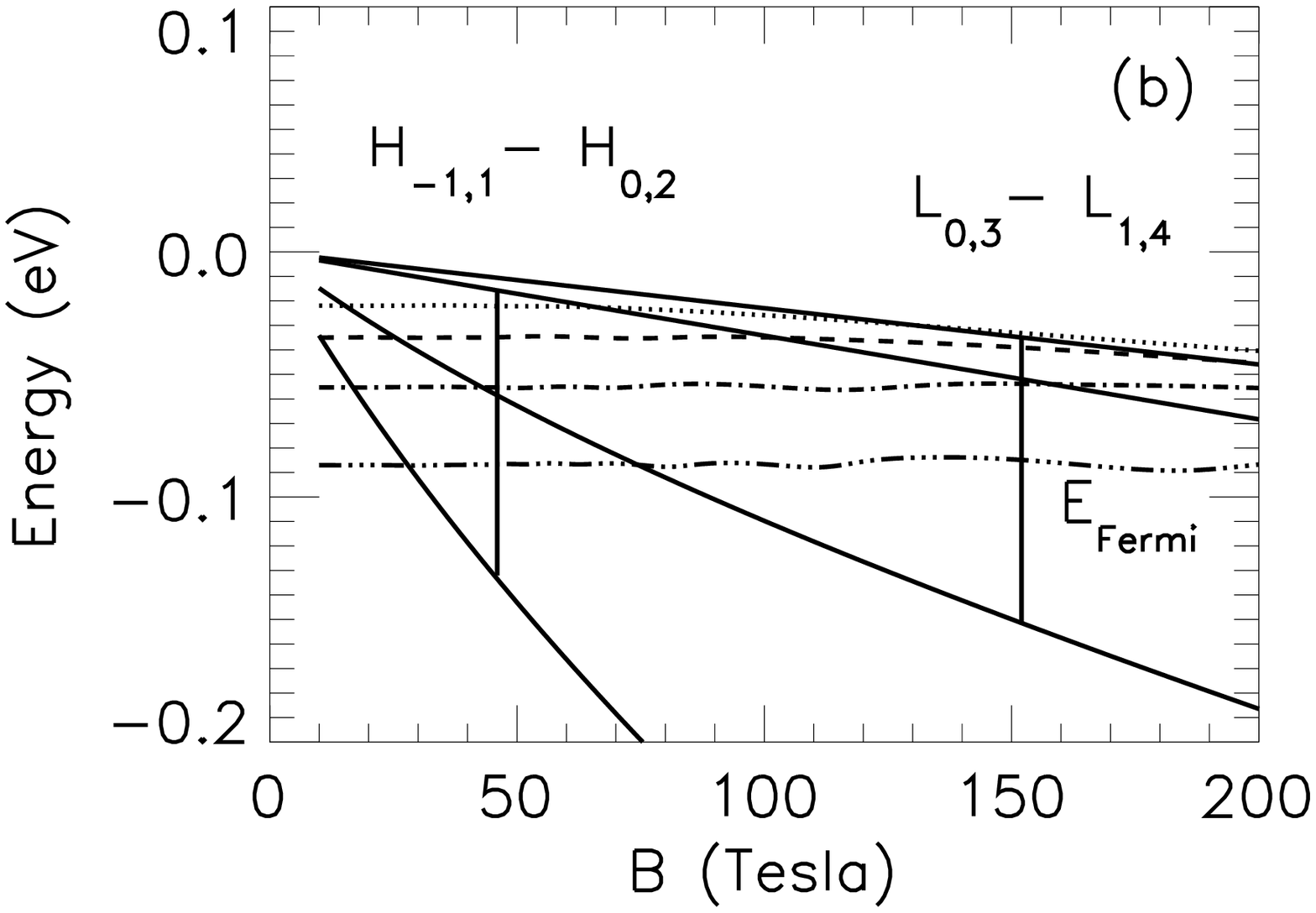}
\caption{Experimental and theoretical h-active CR for p-type
InAs (a). Theoretical CR curves, from bottom to top, assume
hole densities of $5 \times 10^{18}$, $10^{19}$, $2 \times 10^{19}$,
and $4 \times 10^{19} \ \mbox{cm}^{-3}$.
In (b), Landau levels involved in observed CR peaks are
shown along with Fermi levels corresponding to theoretical
curves in (a).}
\label{fig3}
\end{figure}


\section{p-type results}

CR measurements have been performed on p-doped InAs in h-actively
polarized light with $\hbar \omega = 0.117 \ \mbox{eV}$ at a
temperature of 27.5 K. Experimental and theoretical CR spectra are
shown in Fig.~\ref{fig3}(a).
The experimental curve is the negative of the transmission and is 
offset for clarity and plotted in arbitrary units. 
For $B > 30 \ \mbox{T}$, the Fermi energy is such that only the first
two Landau subbands are occupied. Transitions from these levels are
responsible for the CR peaks labeled 1 and 2 in Fig.~\ref{fig3}(a).
The downward sloping CR absorption seen for $B < 30 \ \mbox{T}$ is
due to higher lying Landau levels which become populated at low fields. 
Only the $k = 0$ Landau levels responsible for CR peaks 1 and 2 are
plotted in Fig.~\ref{fig3}(b). Resonant transitions at 0.117 eV
are indicated by vertical lines. The CR peak, 2, near
40 T is due to transitions between the spin-down ground-state heavy
hole Landau subband, $H_{-1,1}$ with energy $E_{-1}(k_z)$ and the
heavy-hole subband, $H_{0,2}$ with energy $E_{0}(k_z)$.
The CR peak, 1, seen at higher fields, is a spin-down light-hole
transition between the $L_{0,3}$ and $L_{1,4}$ subbands. The two 
CR peaks are seen to be asymmetric about their respective resonance
fields. While the theoretical light hole peak does not fit the 
experimental peak position, by varying the Luttinger parameter
$\gamma_1$, the peak position can be brought into better agreement with
experiment. This shows that CR can be used to determine band parameters
and effective masses. The relative strengths of the heavy and
light-hole CR peaks is sensitive to the itinerant hole density and 
can be used to determine the hole density. By comparing theoretical
and experimental curves in Fig.~\ref{fig3}(a), we see that the 
itinerant hole concentration is around
$2 \times 10^{19} \ \mbox{cm}^{-3}$. From Fig.~\ref{fig3}(a), we
can rule out $n < 10^{19} \mbox{cm}^{-3}$ and
$n > 4 \times 10^{19} \mbox{cm}^{-3}$. We estimate that an error
in the hole density of around 25\% should be achievable at these 
densities. 

\begin{figure}[tbp]
\includegraphics[scale=0.42]{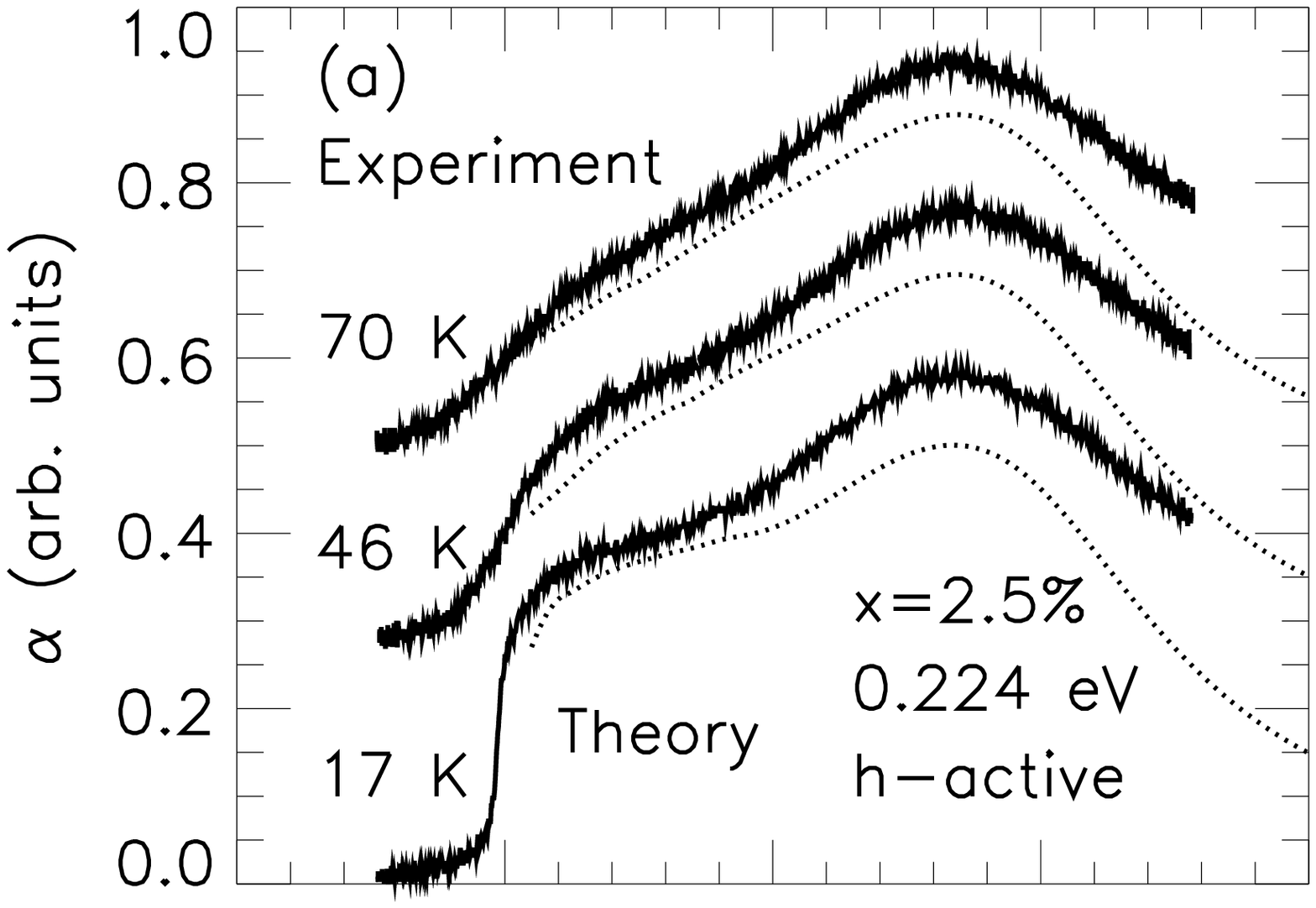}
\includegraphics[scale=0.42]{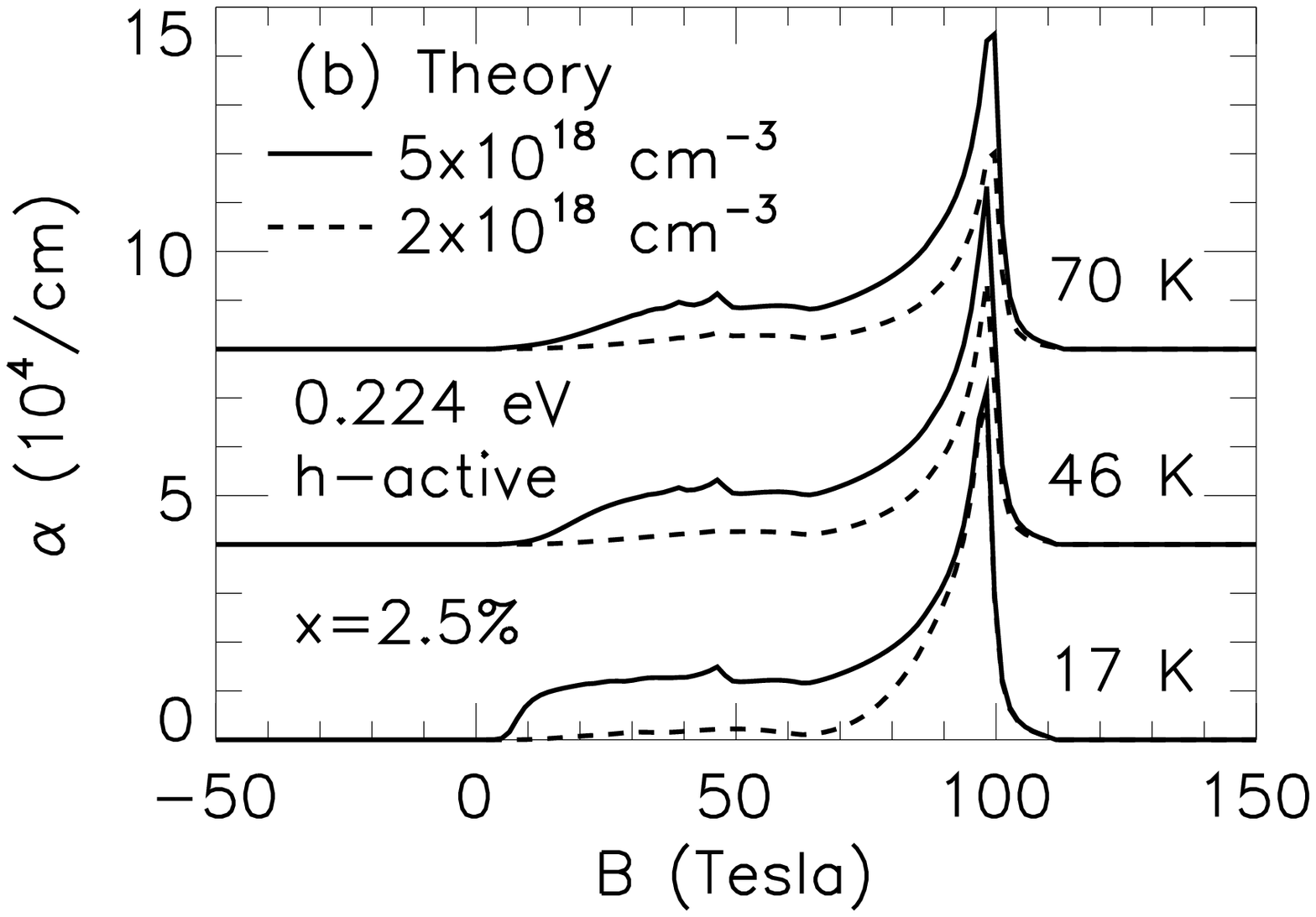}
\caption{CRs for p-type In$_{0.975}$Mn$_{0.025}$As.
Experimental and theoretical CRs are shown in (a) for several
temperatures. In (b), theoretical CRs are shown for two hole densities.}
\label{fig4}
\end{figure}

We next consider the effects of Mn doping on CR in InAs.
We performed CR on p-type In$_{0.975}$Mn$_{0.025}$As at temperatures
of 17, 46, and 70 K in h-active circularly polarized light with
$\hbar \omega = 0.224 \ \mbox{eV}$. It is more convenient to perform
CR at higher photon energies, and hence higher resonant fields,
in order to avoid the low field tails seen in Fig.~\ref{fig3}(a) for
$B < 30 \ \mbox{T}$. This allows us to better examine the lineshapes.
In Fig.~\ref{fig4}(a), the experimental CR is shown as a function of
the magnetic field for the three temperatures and the theoretical
CR spectra are shown as dotted lines. In the theoretical curves, the
FWHM linewidths are taken to be 120 meV and the hole concentration is
taken to be $p = 5 \times 10^{18}\ \mbox{cm}^{-3}$. A heavy hole CR
transition is seen at a resonant field near 80 Tesla. The resonant
field is insensitive to temperature and the lineshape is strongly
asymmetric with a broad tail at low fields. The width of the
low field tail depends on the hole conentration as seen in
Fig.~\ref{fig4}(b) where theoretical CR are computed at two different
hole condentrations assuming a FWHM of 4 meV.
The low field tail is thus a sensitive probe of the hole density.
The sharpness of the low field cutoff depends on temperature and
can be attributed to the sharpness of the Fermi distribution at low
temperatures. In this sample, our estimate of
the hole concentration not all that accurate, but we can see the hole
concentration is greater than $2 \times 10^{18} \mbox{cm}^{-3}$
based on the lack of a low field tail in the latter case.


\section{Conclusions}

We have developed a theory for electronic and magneto-optical
properties in narrow gap InMnAs films and superlattices in ultrahigh
magnetic fields and have shown the CR can be a valuable tool for
determining the densities of itinerant carriers in these systems.
In n-type InMnAs films and superlattices, we found that the e-active
CR peak field is pinned at low electron densities and then begins to
shift rapidly to higher fields above a critical electron concentration
allowing the electron density to be accurately determined.
In p-type materials, determination of hole densities from CR 
measurements is more involved.
In p-type InMnAs, we observed two h-active CR peaks due to heavy and
light holes. The lineshapes were found to
depend on temperature and line broadening. Analyzing
CR lineshapes in p-films and superlatices can help determine hole
densities.

We gratefully acknowledge support from NSF DMR-9817828, 
DMR-0134058 (CAREER), and  DARPA MDA972-00-1-0034 (SPINS).


\end{document}